# Single-Pass Laser Frequency Conversion to 780.2 nm and 852.3 nm Based on PPMgO:LN Bulk Crystals and Diode-Laser-Seeded Fiber Amplifiers

Kong Zhang [1], Jun He [1,2] and Junmin Wang [1,2,*]

[1] State Key Laboratory of Quantum Optics and Quantum Optics Devices, and Institute of Opto-Electronics, Shanxi University, Taiyuan 030006, Shanxi, China
[2] Collaborative Innovation Center of Extreme Optics of the Ministry of Education and Shanxi Province, Shanxi University, Taiyuan 030006, Shanxi, China
* Correspondence: wwjjmm@sxu.edu.cn

**Abstract:** We report the preparation of a 780.2 nm and 852.3 nm laser device based on single-pass periodically poled magnesium-oxide-doped lithium niobate (PPMgO:LN) bulk crystals and diode-laser-seeded fiber amplifiers. First, a single-frequency continuously tunable 780.2 nm laser of more than 600 mW from second-harmonic generation (SHG) by a 1560.5 nm laser can be achieved. Then, a 250 mW light at 852.3 nm is generated and achieves an overall conversion efficiency of 4.1% from sum-frequency generation (SFG) by mixing the 1560.5 nm and 1878.0 nm lasers. The continuously tunable range of 780.2 nm and 852.3 nm are at least 6.8 GHz and 9.2 GHz. By employing this laser system, we can conveniently perform laser cooling, trapping and manipulating both rubidium (Rb) and cesium (Cs) atoms simultaneously. This system has promising applications in a cold atoms Rb-Cs two-component interferemeter and in the formation of the RbCs dimer by the photoassociation of cold Rb and Cs atoms confined in a magneto-optical trap.

## 1. Introduction

The special structure of alkali metal atoms is the foundation of precision spectra, laser cooling and trapping of atoms, atom interferometers, atomic frequency standards, etc. Amongst these atoms, rubidium (Rb) and cesium (Cs) have been studied in the greatest detail. The $5S_{1/2}$–$5P_{3/2}$ transition ($D_2$ line) of rubidium atoms corresponds to 780 nm, and the $6S_{1/2}$–$6P_{3/2}$ transition ($D_2$ line) of cesium atoms corresponds to 852 nm. Therefore, the 780 nm and 852 nm lasers are important. High-power single-frequency 780 nm and 852 nm lasers can be employed for laser cooling and trapping [1–4], atomic coherent control [5,6], atomic interferometers [7,8], and quantum frequency standards [9,10]. In addition, 1.5 μm squeezed light field and quantum-entangled light fields can be prepared from an optical parametric oscillator or optical parametric amplifier pumped by a single-frequency 780 nm laser and has important applications in continuous-variable quantum communication [11], gravitational wave detection [12] and so on. The common means to produce high-power 780 nm and 852 nm laser beams are semiconductor lasers with tapered amplifiers, a master-oscillator power amplifier (MOPA) and Ti:sapphire lasers. However, these systems can only operate in a relatively quiet and clean laboratory environment. In this context, the disadvantages of semiconductor MOPA systems and Ti:sapphire lasers, such as high cost, large size, and sensitivity to vibration and temperature fluctuations, are obvious.

On the other hand, telecom-band lasers have attracted increasing attention in recent years for their extensive and important applications, such as integrated optical devices, all-optical networks, multiwavelength channels [13] and so on. The 1560 nm erbium-doped fiber laser (EDFL) and erbium-doped fiber amplifier (EDFA) are more stable and durable and have been commercialized. These systems can provide a more powerful output than solid-state lasers, and a large number of experiments have been carried out with this system. They even achieved frequency doubling at 780 nm [14] in the plane. In 2015 [15], our group also used fiber lasers and amplifiers to produce a 637.2 nm red laser from single-pass sum-frequency generation (SFG) with two infrared lasers at 1560.5 nm and 1076.9 nm in periodically poled magnesium-oxide-doped lithium niobate (PPMgO:LN) bulk crystal.

Compared with the previous structures, a system that was composed of fiber lasers and amplifiers, and quasi-phase-matched (QPM) non-linear materials, is more remarkable. Therefore, it can be applied in many fields. First, a 780 nm laser can be produced from second-harmonic generation (SHG) with this system, and we have performed these experiments previously [16–18]. The 852 nm laser can be produced from SFG, and there are many research groups working on these experiments. In 2017 [19], Diboune et al. reported mulyi-line operation of all fiber laser system at 780 nm and 852 nm based on waveguides, and they used this system for interferometry of cesium and rubidium atoms. In 2018 [20], Antoni-Micollier et al. demonstrated a watt-level narrow-linewidth fibered laser source at 852 nm by a periodically poled lithium niobate waveguide. The lasers are confined to a very small space in the waveguide, which makes the waveguide a good match with the lasers; however, there will be dispersion and incomplete matching in the bulk crystal. Nonetheless, the bulk crystal can hold more power input, which means that we can obtain a higher power output with a bulk crystal using a high-power injection.

In our experiment, the 780 nm and 852 nm lasers are produced from single-pass SHG and SFG with the 1560 nm and 1878 nm lasers in the PPMgO:LN bulk crystal. Two fundamental lasers are boosted by the corresponding fiber amplifiers, an EDFA for the 1560 nm laser and a thulium-doped fiber amplifier (TDFA) for the 1878 nm laser. The system is particularly robust with single-pass configuration and diode-laser-seeded fiber amplifiers. Therefore, it can work in harsh environments outside a laboratory. It can also produce both 780 nm and 852 nm lasers, so we could conveniently perform laser cooling and trapping, as well as manipulating both Rb and Cs atoms simultaneously, it is useful in the Rb-Cs biatomic interferometer and cold RbCs dimer experiments. Most importantly, the spots produced by this system have good beam quality and a wide tuning range, which make it a scheme that is more promising for application. Considering all of the advantages, the biatomic interferometer and cold RbCs dimer experiments can be performed in unstable environments, such as an airborne system and a system in the space station.

## 2. Experimental Setup

The experimental scheme is shown in Figure 1. The first master-oscillator power fiber amplifier (MOPFA) consists of a compact distributed feedback (DFB) diode laser at 1560.5 nm and a 10 W EDFA, and another MOPFA consists of a compact DFB at 1878.0 nm and a 2 W TDFA. The optical isolator is used to restrain the laser feedback, thus ensuring the stability of EDFA and TDFA. The half-wave plate ($\lambda/2$) and the polarization beam splitter (PBS) cube are used to control the power of the 1560.5 nm and 1878.0 nm lasers, while also transforming the polarization of the fundamental lasers to the s polarization to meet the requirement of the sum-frequency process. Then, a 50 mm-long PPMgO:LN crystal (HC Photonics; Taiwan; the thickness is 0.5 mm; poling period of 23.4 μm; type 0 matching; both ends of the crystal have flat surfaces with anti-reflection coatings for the fundamental and sum-frequency laser, and the residual reflectivity R < 0.2%) is used as the SFG crystal. Additionally, the crystal is placed in a homemade oven, which is made of red copper and stabilized at a precise temperature using a temperature controller (Newport Corp., California, America, Model 350B). We can achieve optimized phase matching by adjusting the temperature of the crystal. Matching is an important factor in a single-pass configuration [21]. We choose f = 100 mm and f = 75 mm with anti-reflection coating for both fundamental and sum-frequency lasers as the matching lens. After the laser passes through the crystal, a 75 mm lens is used to collimate the output laser. Subsequently, the 852.3 nm laser is divided into two beams by PBS: one beam passes through Cs atomic vapor cells and another beam passes through the Fabry–Perot cavity to monitor its frequency tuning range.

Before the SFG, we divide the 1560.5 nm laser into two beams: one beam for the above experiment and another beam for SHG. Single-pass cascaded 25 mm-long PPMgO:LN crystals (HC Photonics; Taiwan; the thickness is 1.0 mm; poling period of 19.48 μm; type 0 matching; both ends of the crystals have flat surfaces with anti-reflection coatings for the fundamental and doubled laser; and the residual reflectivity R < 0.2%) are used as the SHG crystals. We chose f = 50 mm with an anti-reflection coating for both fundamental and second-harmonic lasers as matching lensse. The waist spot radius was ~28 μm, which is very close to the value determined by the optimum B–K focus factor ξ = 2.84 [22].

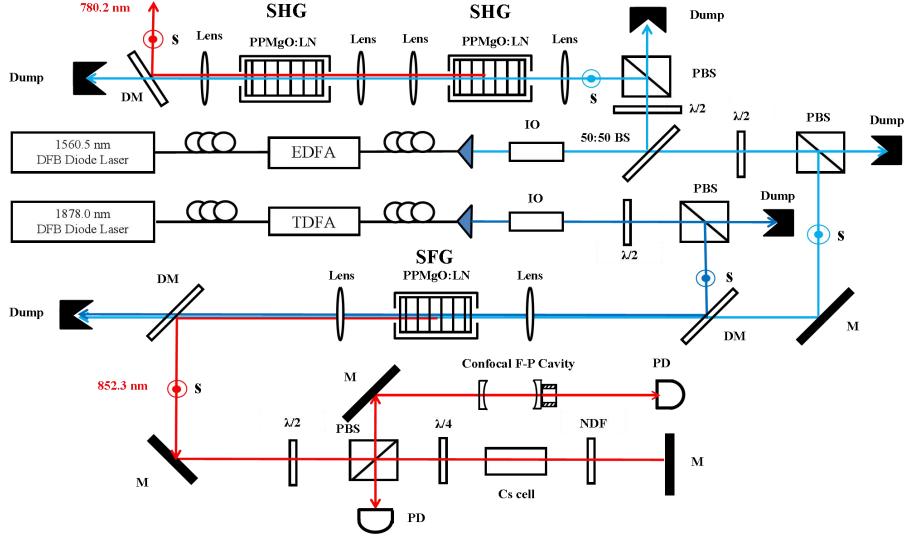

**Figure 1.** Schematic diagram of the single-pass sum-frequency generation (SFG) system; 1560.5 nm and 1878.0 nm lasers are frequency summed in PPMgO:LN to produce 852.3 nm laser. Temperature controllers for three PPMgO:LN crystals are not shown in this figure. DFB: distributed feedback diode laser; EDFA: erbium-doped fiber amplifier; TDFA: thulium-doped fiber amplifier; OI: optical isolator; λ/2: half-wave plate; λ/4: quarter-wave plate; BS: beam splitter plate; PBS: polarization beam splitter cube; DM: dichromatic mirror; NDF: neutral density filters; s: s polarization.

## 3. Experimental Results and Discussion

### 3.1. Single-Pass Second-Harmonic Generation (SHG) to 780.2 nm

First, on the basis of our previous experiment, we find the optimized phase matching by adjusting the temperature of the PPMgO:LN crystal in a single-pass SHG experiment. When the power of the fundamental wave 1560.5 nm is 5.27 W, the best phase matching of the PPMgO:LN-crystal is found at a temperature of 81.8 °C. The results are shown in Figure 2.

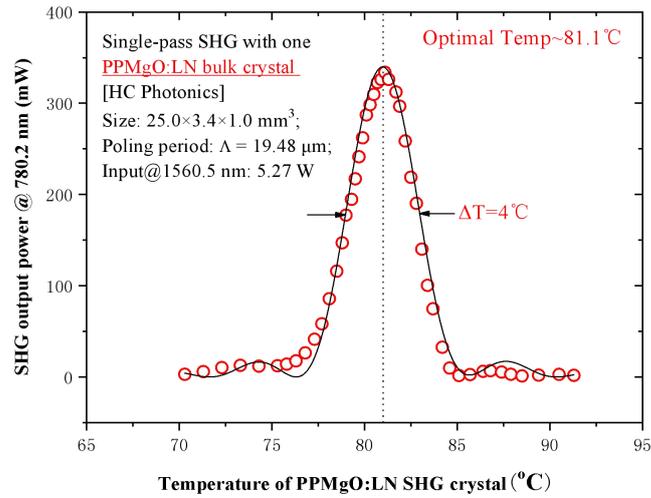

**Figure 2.** Temperature dependence of 780.2 nm laser power. Circles represent the experimental data. The optimal quasi-phase-matched temperature is 81.1 °C.

The single-pass one PPMgO:LN crystal can produce 334 mW for the 780.2 nm laser, and the efficiency is 6.3%; single-pass two PPMgO:LN crystals can produce 634 mW for the 780.2 nm laser, and the efficiency is 12.0%. These results are shown in Figure 3. If we increase the injection, an additional 780.2 nm laser can be produced.

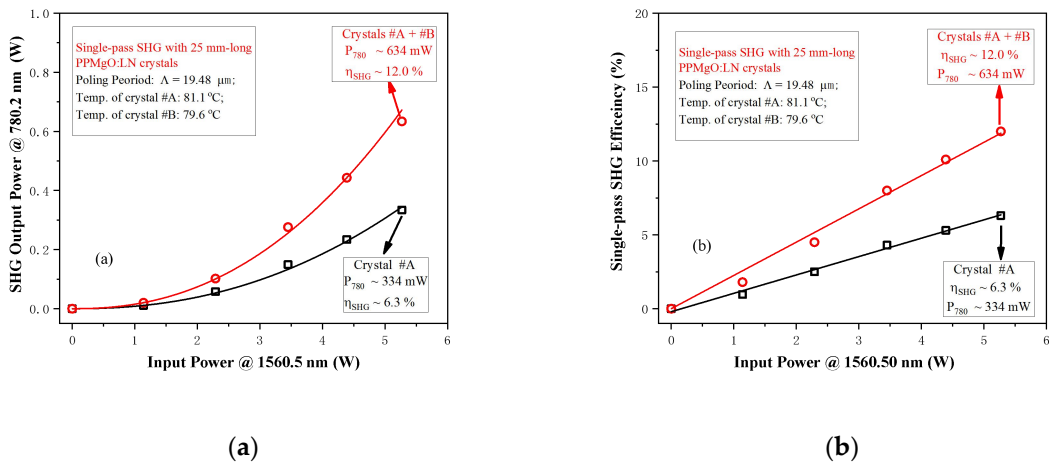

(**a**) (**b**)

**Figure 3.** Experimental data for single-pass second-harmonic generation (SHG) output (**a**) and the frequency doubling efficiency (**b**). The squares represent the case using one PPMgO:LN bulk crystal, and the circles represent the case using two cascaded PPMgO:LN bulk crystals.

The transverse beam quality of the 780.2 nm laser is evaluated by using the beam quality factor parameter ($M^2$) in the two orthogonal transverse directions $X$ and $Y$. Figure 4 shows the $1/e^2$ beam radius versus the axial position Z after passing through a plano-convex lens with a focal length of 80 mm. Fitting of the experimental data gives $M_X^2 = 1.05$ and $M_Y^2 = 1.11$.

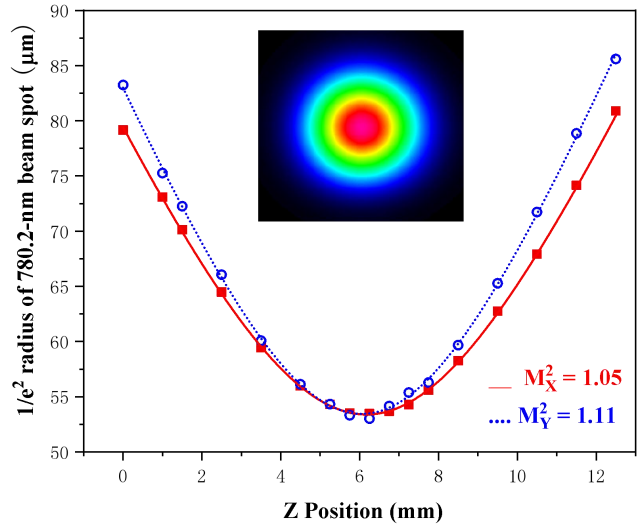

**Figure 4.** Beam quality factor $M^2$ of the SHG laser at 780.2 nm. Typical values for the horizontal and vertical directions are $M_X^2$ = 1.05 and $M_Y^2$ = 1.11. Inset shows the typical intensity profiles of the SHG laser beam spot.

We evaluate the frequency tunability of the 780.2 nm output by measuring the absorption spectra of Rb atomic vapor cells, while the 1560.5 nm fundamental laser frequency is scanned linearly. Figure 5 shows the Doppler-broadened absorption spectra of the $5S_{1/2}$–$5P_{3/2}$ transition ($D_2$ line) for $^{87}$Rb and $^{85}$Rb atoms. These spectra indicate that the continuously tunable range of the doubled laser at 780.2 nm is at least 6.8 GHz.

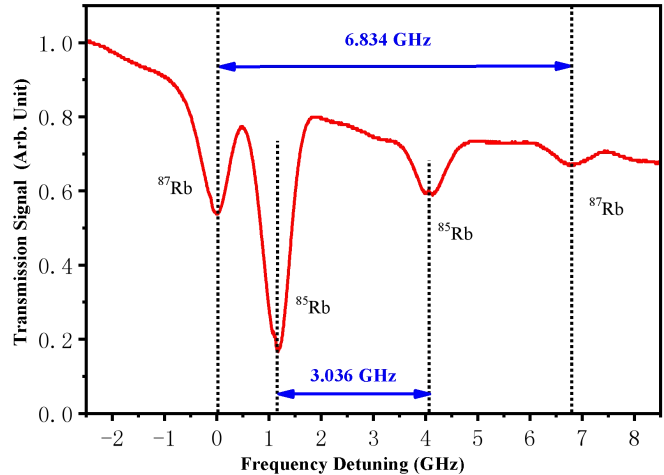

**Figure 5.** Doppler-broadened absorption spectra of $5S_{1/2}$–$5P_{3/2}$ transition ($D_2$ line) of $^{87}$Rb and $^{85}$Rb atoms by using SHG laser at 780.2 nm, while the 1560.5 nm fundamental-wave laser is frequency-scanned linearly. Clearly the continuously tunable range of the 780.2 nm laser frequency is more than 6.8 GHz.

*3.2. Single-Pass Sum-Frequency Generation (SFG) to 852.3 nm*

Then, in a single-pass SFG experiment, we find the optimized phase matching by adjusting the temperature of the PPMgO:LN crystal. When the power of fundamental-wave 1560.5 nm and 1878.0 nm lasers are both 1.0 W, the best phase matching of the PPMgO:LN-crystal is achieved at a temperature of 80.6 °C. The results are shown in Figure 6.

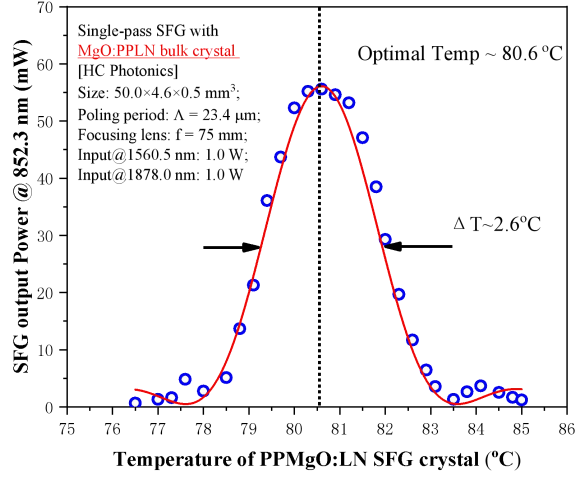

**Figure 6.** Temperature dependence of 852.3 nm laser power. Circles represent the experimental data. The optimal quasi-phase-matched temperature is 80.6 °C.

The results of SFG output power and efficiency are shown in Figure 7. The power of the 1878.0 nm and 1560.5 nm lasers are 1.77 W and 4.94 W, respectively; 276 mW of the 852.3 nm laser can be obtained, and the corresponding optical–optical conversion efficiency is 4.1%. Fitting to the data of the linear region yields a nonlinear conversion efficiency of 0.62% (W cm)$^{-1}$. The SFG nonlinear conversion efficiency can be written as [23]

$$\eta = \frac{p_3}{p_1 p_2 l} = \frac{8\omega_0^3 d_{eff}^2}{\pi \varepsilon_0 c^4 n_0 n_3} \left[ \frac{(1-\delta^2)(1-\gamma^2)}{(1+\delta\gamma)} \right] h(\mu,\zeta) \sin^2(\pi\Delta)$$

where $p_i$ (i = 1, 2, 3) represent the laser power of 1560.5, 1878.0, and 852.3 nm, respectively. $n_o = (n_1 + n_2)/2$, $\omega_0 = (\omega_1 + \omega_2)/2$, $\omega_i$ are the frequency of the corresponding laser, and $d_{eff}$ is the effective nonlinear coefficient. $\delta = 1 - 2\omega_1/(\omega_1 + \omega_2)$ and $\gamma = 1 - 2n_1/(n_1 + n_2)$, $h(\mu,\zeta)$ indicates the B–K focusing factor, which is related to the focusing parameter $\zeta$. The last term is the correct for the crystal's non-ideal grating duty cycle.

In addition, the spatial mode matching of the two fundamental beams within the crystal is the most important factor. The optimal SFG can be obtained when the confocal parameters of these two fundamental beams are equal and the focusing parameter ξ = 2.84. We also use an f = 100 mm lens to match the crystal, and the corresponding optical–optical conversion efficiency is 3.7%. From the experimental results, we can see that when the lens is 75 mm, it can yield a high-power 852.3 nm laser. According to the optimal B–K focus factor ξ = 2.84, the optimal waist spot radii of 1560.5 nm and 1878.0 nm lasers are 45.3 μm and 49.7 μm, respectively. When using a matching lens with f = 75 mm, the waist spot radii of 1560.5 nm and 1878.0 nm lasers are 41.6 μm and 46.5 μm, respectively. The radii for f = 75 mm are closer to the optimal waist spot radius than those for f = 100 mm. In a single-pass sum-frequency experiment, the matching lens is not the only factor that influences efficiency, but the overlapping of the two fundamental-wave lasers in the crystal is also important. Although an achromatic doublet can make 1560.5 nm and 1878.0 nm lasers at one point in the crystal, but it is not easy to customize. Therefore, we will also study the method of separate focusing to make the fundamental waves match well, including size and position.

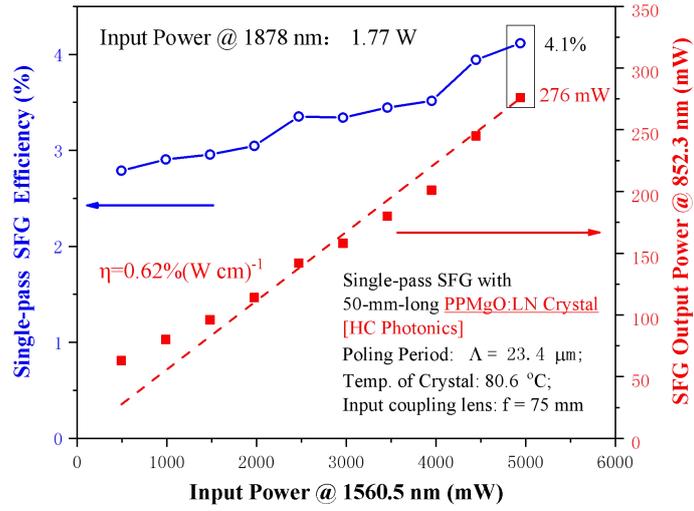

**Figure 7.** Experimental data for single-pass SFG using a PPMgO:LN crystal with matching lens f = 75 mm. The solid squares represent output power of 852.3 nm, while the circles represent the corresponding SFG conversion efficiency.

The transverse beam quality of 852.3 nm laser is evaluated by using the beam quality factor parameter ($M^2$) in the two orthogonal transverse directions $X$ and $Y$. Figure 8 shows the $1/e^2$ beam radius versus the axial position Z using a plano-convex lens with a focal length of 50 mm. Fitting of the experimental data gives $M_X^2 = 1.07$ and $M_Y^2 = 1.13$.

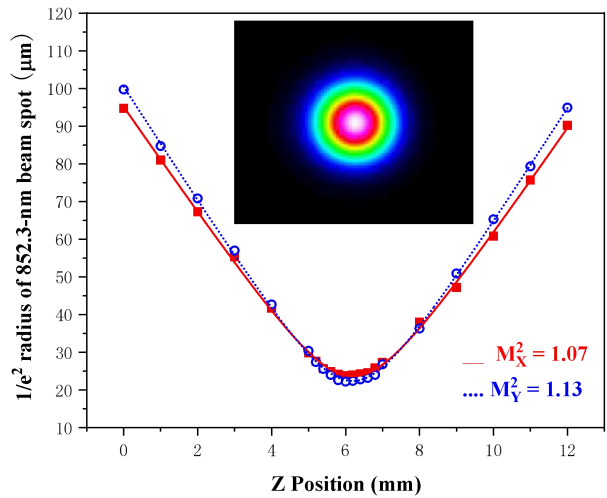

**Figure 8.** Beam quality factor $M^2$ of the SFG laser beam at 852.3 nm. Typical values for the horizontal and vertical directions are $M_X^2 = 1.07$ and $M_Y^2 = 1.13$. Inset shows the typical intensity profiles of the SFG laser beam spot.

When we fix the 1878.0 nm laser and scan the 1560.5 nm laser, we can obtain cesium absorption spectra by Cs atomic vapor cells. Then, we analyze the transmitted signal of a confocal Fabry–Perot cavity with a free spectral range (FSR) of 750 MHz while the 852.3 nm laser frequency is linearly scanned. Figure 9 shows a typical result, where the 852 nm laser can be tuned across more than 12 FSRs, which means that the continuously tunable range is at least 9.2 GHz. The scanning range is mainly limited by the DFB. If the scanning range of the fundamental frequency laser is sufficiently large, the sum-frequency 852.3 nm laser can also be scanned following. This is another advantage compared with cavity-enhanced scheme.

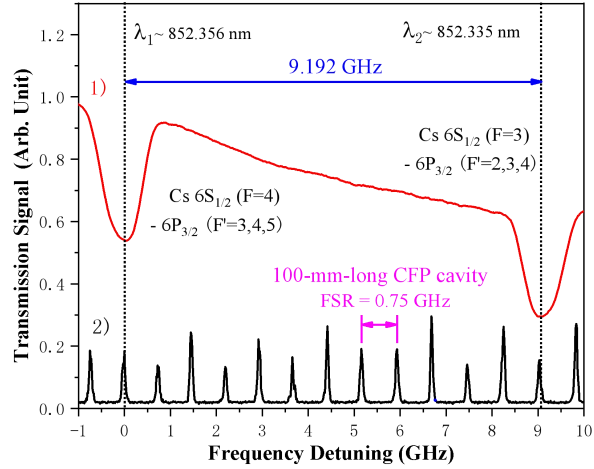

**Figure 9.** The trace (1) shows the $D_2$ line absorption spectra of $6S_{1/2}$–$6P_{3/2}$ transition of cesium atoms by using SFG laser at 852.3 nm, while the 1560.5 nm fundamental-wave laser is frequency-scanned linearly, but the 1878.0 nm fundamental-wave laser's frequency is fixed. Frequency scanned 852.3 nm laser beam is monitored using a confocal Fabry–Perot cavity with free spectral range (FSR) of 0.75 GHz as shown by the trace (2). Clearly the continuously tunable range of the 852.3 nm laser's frequency is more than 9.2 GHz.

## 4. Conclusions

We have implemented a simple, robust and efficient scheme that combines single-pass SHG at 780.2 nm with single-pass SFG at 852.3 nm based on diode-laser-seeded fiber amplifiers. A power of more than 630 mW of the 780 nm single-frequency continuous-wave laser is demonstrated through two PPMgO:LN cascaded bulk crystals, and the SHG efficiency obtained is 12.0%. At the same time, a power of more than 270 mW of the 852 nm single-frequency continuous-wave laser is reported with the PPMgO:LN bulk crystal, and the SFG efficiency is 4.1%. In addition, the continuously tunable ranges of the 780 nm and 852 nm laser beams are at least 6.8 GHz and 9.2 GHz. Although, the size of our device is large, its mechnical stability can be further improved by careful design with a much more compact structure.

Most importantly, the output lasers have a high beam quality, which cannot be addressed by using a diode laser combined with a semiconductor tapered amplifier. In addition, we can confine Rb atoms by using an 852 nm optical dipole trap on the basis of a 780 nm magneto-optical trap. Previously, we required 852 nm and 780 nm diode lasers to perform such an experiment, but our current system can produce both lasers at high power. Finally, the system has great potential for application in the formation of RbCs dimers via photoassociation of cold Rb and Cs atoms and an Rb–Cs two-component atomic interferemeter.

**Author Contributions:** The experiment and data were completed by kong Zhang. Jun He undertaken guidance during the experiment. And Junmin Wang's contributions included coordination, guidance and analysis in the experiment.

**Funding:** This research was funded by [the National Key R&D Program of China] grant number [2017YFA0304502], [the National Natural Science Foundation of China] grant number [11774210 and 61875111], and [the Shanxi Provincial 1331 Project for Key Subject Construction] grant number [201542033].